\documentclass[superscriptaddress,aps,prb,twocolumn,noshowpacs]{revtex4-2}
\usepackage{graphicx}
\usepackage{epstopdf}
\usepackage{dcolumn}
\usepackage{bm}
\usepackage{color}
\usepackage{amsmath}
\usepackage{ulem}

\begin{document}

\title{Superconducting FeSe monolayer with milli-electron volt Fermi energy}

\author{Wantong Huang}
\affiliation{State Key Laboratory of Low-Dimensional Quantum Physics, Department of Physics, Tsinghua University, Beijing 100084, China.}
\affiliation{Institute of Flexible Electronics Technology of THU, Zhejiang}

\author{Haicheng Lin}
\affiliation{State Key Laboratory of Low-Dimensional Quantum Physics, Department of Physics, Tsinghua University, Beijing 100084, China.}

\author{Cheng Zheng}
\affiliation{State Key Laboratory of Low-Dimensional Quantum Physics, Department of Physics, Tsinghua University, Beijing 100084, China.}

\author{Yuguo Yin}
\affiliation{State Key Laboratory of Low-Dimensional Quantum Physics, Department of Physics, Tsinghua University, Beijing 100084, China.}

\author{Xi Chen}
\email{xc@mail.tsinghua.edu.cn}
\affiliation{State Key Laboratory of Low-Dimensional Quantum Physics, Department of Physics, Tsinghua University, Beijing 100084, China.}
\affiliation{Frontier Science Center for  Quantum Information, Beijing, China}

\author{Shuai-Hua Ji}
\email{shji@mail.tsinghua.edu.cn}
\affiliation{State Key Laboratory of Low-Dimensional Quantum Physics, Department of Physics, Tsinghua University, Beijing 100084, China.}
\affiliation{Frontier Science Center for  Quantum Information, Beijing, China}
\affiliation{RIKEN Center for Emergent Matter Science (CEMS) - Wako, Saitama 351-0198, Japan}

\date{\today}

\begin{abstract}
Iron selenide (FeSe) is an iron-based superconductor which shows unique properties, including strongly anisotropic superconducting gap, paramagnetism in undoped compound and extremely small Fermi pocket size. In this work, we demonstrate that the sizes of electron and hole pockets in FeSe monolayer become much smaller than those in bulk. The Fermi energy is in the order of a few meV and can be fine-tuned by the thickness of graphene layers underneath. Despite the low carrier density, the FeSe monolayers grown on trilayer or multi-layer graphene are superconducting. The superconducting gap size is sensitive to the Fermi energy of the hole band. Remarkably, the FeSe monolayer provides the opportunity to study the physics in the crossover regime where the Fermi energy and superconducting gap are comparable to each other.

\end{abstract}

\maketitle

\section{Introduction}

New physics often emerges when two energy scales becomes comparable. It is highly desirable to discover materials in such a crossover regime.  Here we show that monolayer iron selenide (FeSe) is an eligible candidate.  FeSe is a compensated semimetal and owns the simplest crystalline structure \cite{Paglione10} among the iron-based superconductors while exhibiting the characteristic tetragonal-orthorhombic structure transition \cite{Wu08} and nematicity \cite{baek2015orbital,PhysRevB.94.201107} in common. 
In the tetragonal phase, the Fermi surface of bulk FeSe consists of ellipsoidal hole pockets around $\Gamma$=(0,0) point and electron pockets around X=($\pi$/$a_{Fe}$,0) points  [Fig. 1(a)]. The hole and electron pockets  are predominately contributed by the $d_{xz}$ and $d_{yz}$ orbitals, respectively. It has been demonstrated by angle-resolved photon-electron spectroscopy \cite{PhysRevX.8.031033,PhysRevB.94.201107,PhysRevB.92.205117,PhysRevB.91.155106}, scanning tunneling spectroscopy (STS) \cite{kasahara2014field,Davis18,PhysRevLett.123.216404}, and transport measurements \cite{kasahara2014field,Uji14,PhysRevB.91.155106,Audouard15} that the pockets are rather shallow with Fermi energy $\epsilon_F$ around 10 meV. Furthermore, the Fermi energy can be tuned via chemical doping, such as S\cite{Hanaguri2018Two}-  and Te-substitutions \cite{Lubashevsky2012Shallow,2014Superconductivity,Rinott2017Tuning}. In Fe$_{1+y}$Se$_x$Te$_{1-x}$,  the Fermi energy of the hole band is reduced from 19 to 6 meV with decreasing excess Fe concentration y  \cite{Rinott2017Tuning}.  The Seebeck coefficient \cite{PhysRevB.83.020504}  sets the upper limit of the Fermi energy of electron band in Fe$_{1+y}$Se$_{0.4}$Te$_{0.6}$ to be $\sim$10 meV.  Conceivably, the pockets can become even smaller in monolayer FeSe because the absence of inter-layer coupling  tends to narrow the energy band and decrease the overlap between electron and hole pockets in energy. In light of this expectation, we have been able to achieve a Fermi energy of a couple of meV in monolayer FeSe, which is almost one order of magnitude lower than that in bulk and  comparable to the superconducting gap. In term of the Uemura plot \cite{Uemura89}, the T$_c$/T$_F$ ratio of FeSe monolayer is higher than most of the unconventional superconductors. Moreover, the in-depth study of monolayer FeSe has it own importance. The transition temperature of monolayer FeSe grown on SrTiO$_3$ is greatly enhanced as manifested by various  investigations \cite{Xue12,Zhou12,Zhou13,Feng13}. The enhancement is mainly interpreted by the strong coupling between FeSe and SrTiO$_3$. To elucidate the mechanism, it is desirable to reveal the intrinsic properties of an almost free-standing FeSe monolayer, which leads us to grow the film on van der Waals substrate. We also show that the charge transfer from the substrate can fine-tune the Fermi energy of the FeSe monolayer.

\section{Experimental Details}
The experiments were performed on a low temperature ultra-high vacuum  (UHV, 1$\times$10$^{-10}$ torr)  scanning tunneling microscope (STM) equipped with molecular beam epitaxy (MBE). The lowest temperature of STM head could reach base temperature of 60 mK with relative high effective electronic temperature of 260 mK in samples \cite{chen19}. To prepare the FeSe monolayer, which has been grown on SrTiO$_3$ \cite{Xue12} and other substrates \cite{PhysRevLett.117.067001, Xue18}, high-purity Fe (99.995$\%$) and Se (99.999$\%$)  were co-deposited onto the n-type 6H-SiC(0001) substrate (nitrogen-doped, resistivity 0.02-0.2 $\Omega\cdot$cm) held at 400$^\circ$C. To reduce the coupling between FeSe and the substrate to the van der Waals type \cite{PhysRevB.84.020503}, the surface of SiC was graphitized in advance by thermal desorption of Si from the topmost layers.  Both bi-layer (BLG) and tri-layer graphene (TLG) can be formed and their relative coverage depends on the heating temperature (1400$^\circ$C$\sim$1450$^\circ$C) and duration time.  The growth of FeSe was carried out under Se-rich condition and monitored by {\it in situ} reflection high-energy electron diffraction. The growth rate was about two monolayers per hour. 

\begin{figure*}[htp]
       \begin{center}
       \includegraphics[width=6.75in]{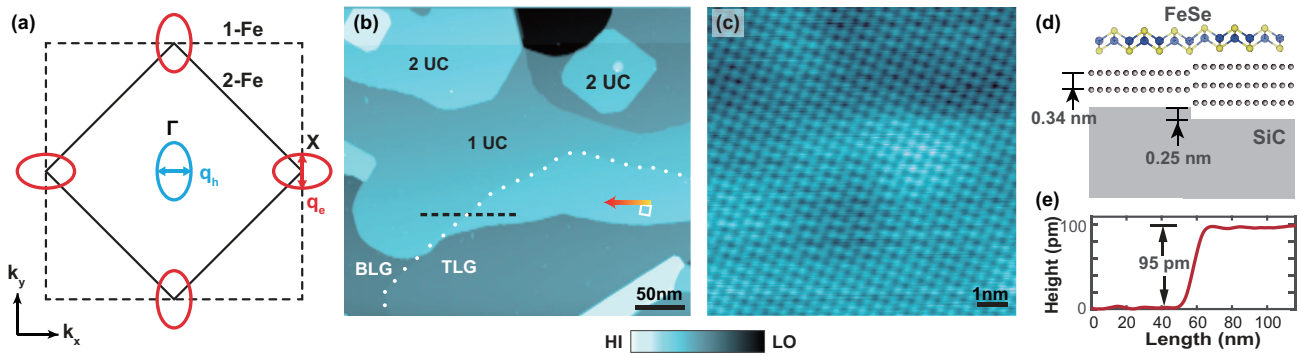}
       \end{center}
       \caption[]{
(a)  Schematic of the Brillouin zone and the Fermi surface. In the nematic phase, the electronic structure should be more properly viewed in the two-Fe Brillouin zone.
(b) Topographic image (350 nm$\times$315 nm) of FeSe islands acquired by using sample bias of $V$=3 V and tunneling current of $I$=20 pA. 1 UC (unit cell) and 2 UC are the areas for monolayer and bilayer FeSe. 
(c)  Atomically resolved STM topography (10 nm$\times$10 nm, 0.1 V, 0.1 nA) of the area marked by the white square in (b). (d) Side view of the  monolayer FeSe across the step between BLG and TLG on adjacent SiC terraces. (e) Topographic profile along the black dashed line in (b).
}
\end{figure*}

The electronic structure of FeSe monolayer was studied by STM and STS. To avoid any contamination, we performed the STM experiments on the films in the same UHV system as MBE. Throughout the experiments, the STM remained at the base temperature. Before imaging, the polycrystalline Pt-Ir alloy tip was modified and calibrated on a clean Ag(111) surface. In the measurement, the d$I$/d$V$ spectra on FeSe films were acquired by the standard lock-in technique with a modulation frequency $f=887$ Hz. 

\section{Results}

Figure 1(b) shows the topographic image of FeSe films on the substrate covered with both BLG and TLG whose boundary is indicated by white dotted line. The lateral size of a film is usually a few hundred nanometers. 
Atomically resolved STM image reveals the top Se atoms with Se-Se distance of 3.75 \AA{} [Fig. 1(c)]. In the nematic phase below $\sim$90 K, FeSe unit cell has two inequivalent Fe-Fe distances: $a_{Fe}$=2.665 \AA{} and $b_{Fe}$=2.655 \AA \cite{PhysRevLett.103.057002}. Such a tiny difference is beyond the resolution of STM and could not be resolved. The lattice of FeSe monolayer is continuous across the border between BLG and TLG. As illustrated in Fig. 1(d), TLG is about 0.9 \AA{} higher than BLG because one more layer of SiC needs to be depleted to form TLG. The 0.9 \AA{} difference also presents in the apparent height of FeSe monolayer (see the profile in Fig. 1(e)). 

\begin{figure}[htp]
       \begin{center}
       \includegraphics[width=3.25in]{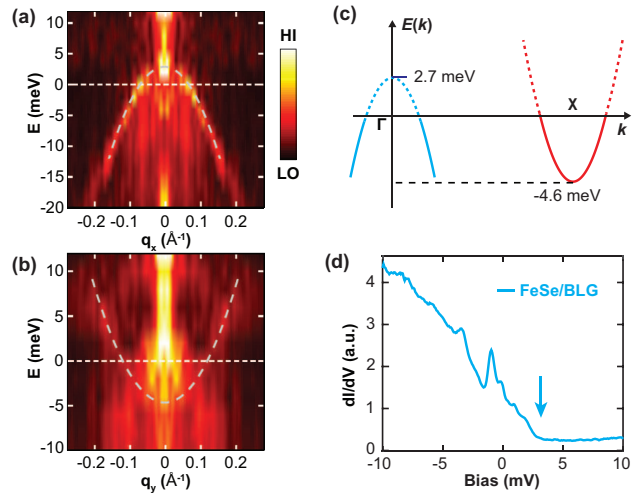}
       \end{center}
       \caption[]{
(a-b) QPI dispersions of the non-superconducting area of FeSe monolayer grown on BLG, obtained by taking line cuts from the Fourier transform of energy-dependent normalized conductance images (Supplementary Fig. S2\cite{SI}) along $q_x$ and $q_y$, respectively. The dispersions are fitted by the dashed curves. The tunneling spectra of the mappings were taken on a grid of 128$\times$128 pixels for a 85 nm$\times$85 nm field of view. Sample bias voltage $V$=20 mV, tunneling current $I$=100 pA, modulation amplitude for the lock-in detection $V_{mod}$=0.2 mV.    
(c) Schematic of the band dispersion around $\Gamma$=(0,0) point and X=($\pi$/a$_{Fe}$,0) point. 
(d) The d$I$/d$V$ spectrum ($V$=10 mV, $I$=0.1 nA, $V_{mod}$=0.1 mV) of monolayer on BLG. The arrow marks the top of hole pocket. The peaks are caused by the quantum confinement in the lateral direction. 
}

\end{figure}

High quality of the FeSe monolayer film enables us to estimate the Fermi energies of the hole and electron bands via the quasi-particle interference (QPI) imaging. QPI visualizes the elastic scattering of electrons on the constant-energy contour by mapping the energy-dependent normalized differential tunneling conductance (Supplementary Fig. S1\cite{SI}) on the surface. Thereby the Fourier transform of QPI provides information of the energy-momentum dispersion.  Such spectroscopic mapping was performed on the FeSe monolayer grown on BLG. The Fourier transform (Supplementary Fig. S2\cite{SI}) exhibits strong in-plane anisotropy because of the orbital selective coherence \cite{Davis18}. The intra-pocket scattering wave vectors $q_e$ and $q_h$ in Fig. 1(a) for the electron and hole pockets can be identified in the Fourier transform pattern. The band dispersions are extracted  [Figs. 2(a-b)] and then fitted with parabolic curves. The fitting gives the effective mass of holes (electrons) of 1.5$\pm$0.1 $m_0$ (2.9$\pm$0.2 $m_0$) along the corresponding directions $k_x$ ($k_y$) in the momentum space, where $m_0$ is the free electron mass.
Notably, the top of the hole band is at 2.7$\pm$0.4 meV and the bottom of the electron band at -4.6$\pm$0.5 meV [Fig. 2(c)], respectively. Such small Fermi energies have exceeded all the previous efforts on FeSe. However, it is still unsatisfactory since the monolayer film grown on BLG shows no signature of superconductivity  [Fig. 2(d)] at the base temperature (monolayer FeSe on graphene has also been confirmed to be non-superconductive at 2.2 K  before\cite{PhysRevB.84.020503}). 

\begin{figure}[htp]
       \begin{center}
       \includegraphics[width=3.25in]{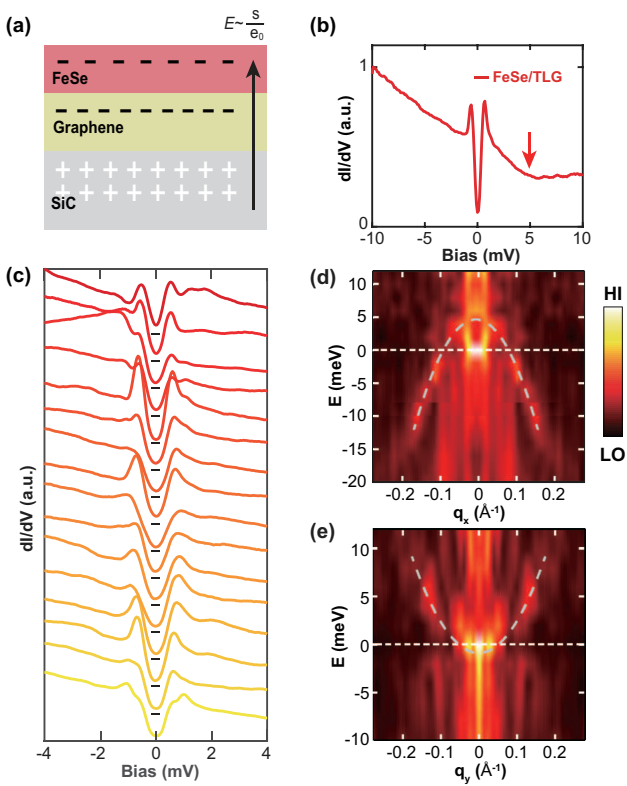}
       \end{center}
       \caption[]{
(a) The dipole layer formed between SiC and FeSe surface.
(b) The d$I$/d$V$ spectrum ($V$=10 mV, $I$=0.1 nA, $V_{mod}$=0.1 mV) in the superconducting area. 
(c) A series of d$I$/d$V$ spectra ($V$=5 mV, $I$=0.1 nA, $V_{mod}$=0.05 mV) measured along the arrow of 50 nm long in Fig. 1(b).
(d-e) QPI dispersions of the superconducting area of FeSe monolayer grown on TLG (Supplementary Fig. S3\cite{SI}).
}

\end{figure}

FeSe monolayer on graphene can become superconducting by hole doping from substrate as shown in Fig. 3(a).  In thermal equilibrium, the chemical potentials of SiC, graphene and FeSe should be aligned as a result of charge transfer. The alignment leads to the formation of a dipole-layer of a few nanometer thick below the surface [Fig. 3(a)]. The charge distribution inside the dipole-layer depends on the detailed structure at the atomic level. For FeSe/graphene/SiC structure, the carrier density and Fermi energy of FeSe are closely related to the thickness of the graphene layers underneath. In this case, the characteristic energy and length scales are 0.1 eV (Supplementary Fig. S4\cite{SI}) and 1 nm, respectively. The corresponding carrier density induced by electric field ($\sim$energy/length) is estimated to be $10^{12}\text{ electrons}/\text{cm}^{2}$ [Fig. 3(a)]. Given the density of states for FeSe monolayer as $10^{15}\text{ electrons}/(\text{eV}\cdot\text{cm}^2)$ (Supplementary section V\cite{SI}), the change in the carrier density of FeSe on different thickness of graphene layers can bring about a shift of Fermi energy in the order of meV. Such a shift is significant in manipulating the electronic properties if the Fermi energy is also in the similar range. 


More specifically, the charge neutral point of TLG moves upward in energy by about 0.1 eV compared to BLG (Supplementary Fig. S4\cite{SI}). Therefore, the FeSe film on TLG should be considerably hole doped. As a result, the FeSe film on TLG in Fig. 1(b) becomes superconducting. The spectrum in Fig. 3(b) shows a typical gap of $\Delta$=0.60 meV at the base temperature.  The line spectra [Fig. 3(c)] taken along the arrow in Fig. 1(b) exhibit certain inhomogeneity in spatial distribution.  

\begin{figure}[htp]
       \begin{center}
       \includegraphics[width=3.25in]{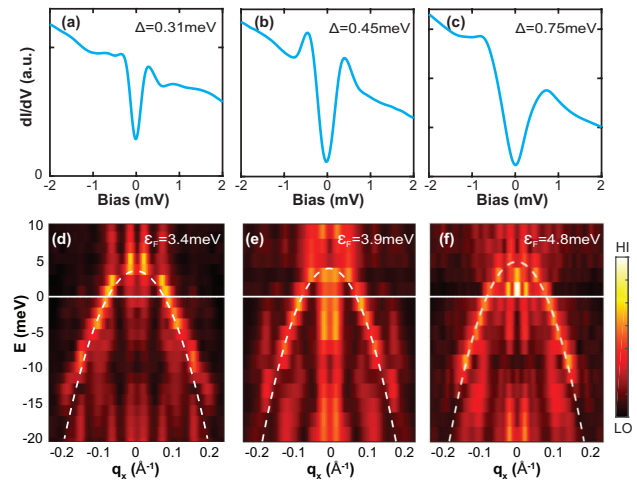}
       \end{center}
       \caption[]{
 (a-c) The average d$I$/d$V$ spectra of FeSe monolayer on TLG. The spectrum in (a) is the average of a line cut taken along a 60 nm long line (64 points evenly distributed along this line). Set point: $V$=2 mV, $I$=100 pA, $V_{mod}$=0.02 mV. The spectrum in (b) is the average of a line cut taken along a 50 nm long line (32 points evenly distributed along this line). Set point: $V$=-4 mV, $I$=100 pA, $V_{mod}$=0.04 mV. The spectrum in (c) is the average of a line cut taken along a 48 nm long line (32 points evenly distributed along this line). Set point: $V$=10 mV, $I$=100 pA, $V_{mod}$=0.1 mV. (d-f) The corresponding QPI dispersions acquired in the same areas as the upper panel (Supplementary Fig. S6\cite{SI}).  
}
\end{figure}

\begin{figure*}[htp]
       \begin{center}
       \includegraphics[width=6.5in]{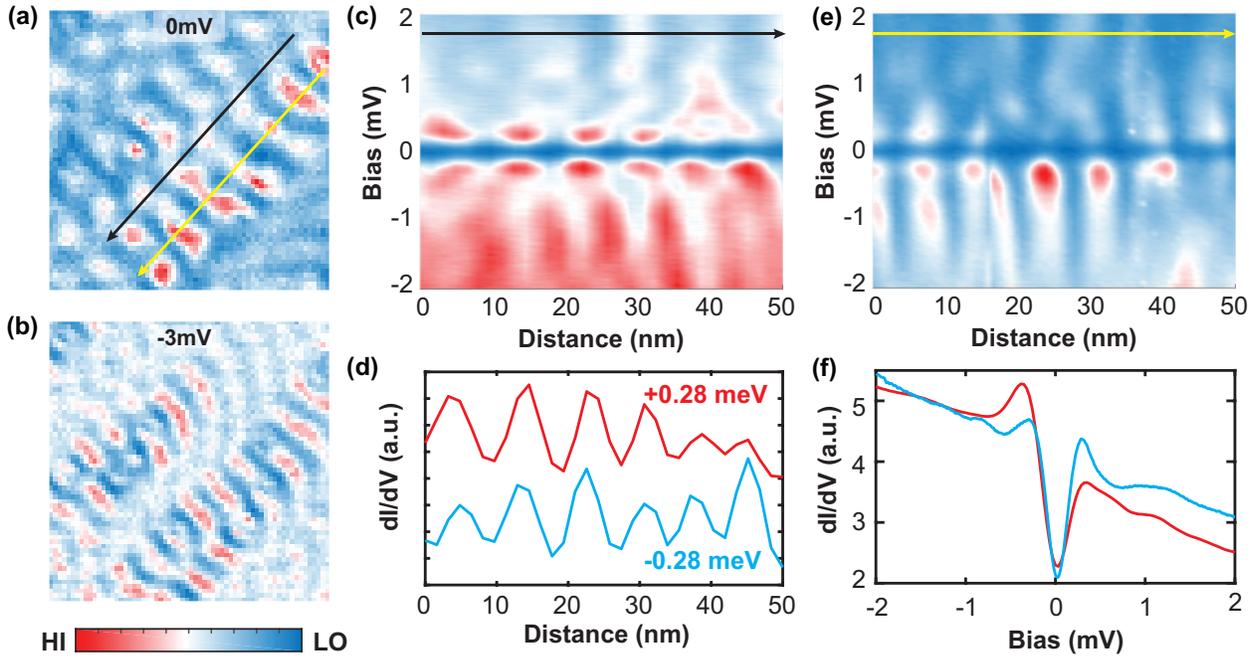}
       \end{center}
       \caption[]{
(a-b) d$I$/d$V$($r$,$E$) mapping measured on a 50 nm $\times$ 50 nm area of FeSe monolayer films. Standing waves are clearly visible. Set point: $V$=-21 mV, $I$=0.1 nA, $V_{mod}$=0.21 mV. Lock-in oscillation amplitude 0.21 mV. 
(c)  A series of d$I$/d$V$ spectra along a black arrow of 50 nm long in (a)(32 points evenly distributed along this line). The averaged gap size is 0.28 meV. Set point: $V$=-5 mV, $I$=100 pA, $V_{mod}$=0.05 mV.
(d) d$I$/d$V$ at 0.28 mV and -0.28 mV along the line.  
(e)  A series of d$I$/d$V$ spectra along a yellow arrow of 50 nm long in (a)(64 points evenly distributed along this line). Set point: $V$=-10 mV, $I$=100 pA, $V_{mod}$=0.1 mV.
(f) The blue and red curves are the averaged d$I$/d$V$ spectra of (c) and (e) respectively. The gap size of blue (red) curve is 0.28 meV (0.35 meV). 
}
\end{figure*}

The hole doping is confirmed by QPI measurement on FeSe monolayer on TLG. Figures 3(d) and 3(e) show the band dispersions extracted from QPI.  The top of the hole band  for FeSe monolayer on TLG [Fig. 3(d)] is at $\sim$4.7$\pm$0.5 meV and 2 meV higher than that for the monolayer on BLG [Fig. 2(a)]. The 2 meV shift comes from the hole doping as expected from the estimation based on chemical potential alignment between film and substrate.  For the electron pocket, the bottom of the band is estimated to be -1.3$\pm$1.5 meV [Fig. 3(e)]. Using previous ARPES data on the Fermi surface anisotropy\cite{PhysRevX.8.031033}, the hole and electron densities are estimated to be 8.9$\times$10$^{12}$ cm$^{-2}$ and 1.5$\times$10$^{13}$ cm$^{-2}$, respectively. Such a low carrier density for a superconductor is rare except in some transition metal dichalcogenide monolayers\cite{fatemi2018electrically,sajadi2018gate} and recently discovered twisted bilayer graphene\cite{Jarillo-Herrero18}.

A series of FeSe monolayer films have been prepared on TLG. Depending on the locations of graphene on SiC, the doping level varies.  On each film, a relatively large uniform area was chosen for STS and QPI measurement.  Figures 4(a-c) display the averaged d$I$/d$V$ spectra of FeSe monolayer all on TLG but with  different doping levels. The gap sizes of the three areas are  0.31 meV, 0.45 meV, and 0.75 meV, respectively. The corresponding Fermi energy $\epsilon_F$ of the hole band in each case can be obtained by fitting the QPI dispersions [Figs. 4(d-f)] and found to be 3.4 meV, 3.9 meV and 4.8 meV, respectively. It is evident that the superconducting gap size $\Delta$ is sensitive to and increases monotonically with the Fermi energy of the hole band [see also Fig. S5].  The increased hole density in the FeSe monolayer enhances the screening effect and hence the superconductivity. This observation may also indicate that the superconductivity of FeSe monolayer is dominated by a single hole band.

All the dispersions in Figs. 3 and 4 were obtained in carefully selected areas to make sure that the Fermi energy of each area is highly uniform.  The Fourier transform of the standing wave generated by the interference of quantum states scattered off point defects and step edges in an area leads to the dispersion. The spatial uniformity of the dispersion in each area is demonstrated by the clearly-defined Fourier transform pattern [Fig. S6]. Inside a superconducting area, all the spectra showing the superconducting gaps  in Figs. 3 and 4 were obtained along  lines away from the defects. In the vicinity of defects, on the other hand,  the spectra reveal large variation. For example, the Bogoliubov quasiparticles around the coherence peaks can present the standing wave pattern [Figs. 5(c-e)] due to the scattering of defects.  The coherence peaks locate at $\pm$0.28 meV and the corresponding wavelengths are 16.78 nm (at +0.28 meV) and 16.12 nm (at -0.28 meV) [Fig. 5(d)], respectively. The line spectra [Fig. 5(e)], taken along the yellow line where the standing wave is stronger, have a relatively larger gap size of 0.35 meV [Fig. 5(f)]. Even more inhomogeneity on the nanometer scale presents  in highly disordered areas, where the superconducting gap  exhibits strong spatial variations [Figs. S7]. Detailed studies are needed to understand the implications of various spectra.  

So far it has been demonstrated that FeSe monolayer is a unique material: The Fermi energy $\epsilon_F$ can be pushed down to the meV range and thus becomes comparable to the superconducting gap $\Delta$. As a result, it is possible to realize novel quantum states in FeSe monolayer, such as BCS-BEC crossover \cite{kasahara2014field}, which need further investigation.  The extremely low Fermi energy is inherent only to the monolayer. For those areas in Fig. 1(b) with bilayer FeSe, the dispersion of the hole pocket corresponds to a Fermi energy of $15\sim20$ meV [Fig. 6(a)], which is close to the bulk counterpart and much larger than that of the FeSe monolayer. The inter-layer coupling gives rise to  one more hole band crossing the Fermi level. Usually the bilayer FeSe is superconducting [Fig. 6(b)]. 
\begin{figure}[htp]
       \begin{center}
       \includegraphics[width=3.25in]{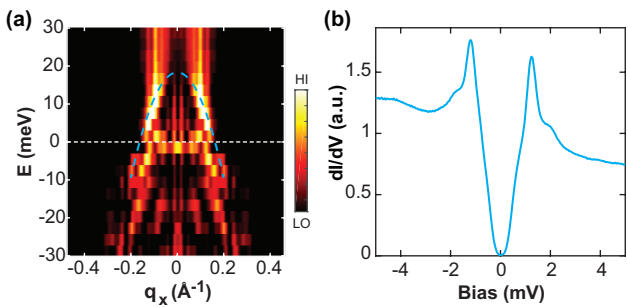}
       \end{center}
       \caption[]{
(a) QPI dispersion of FeSe bilayer. (b) d$I$/d$V$ of superconducting FeSe bilayer. Set point: $V$=5 mV, $I$=100 pA, $V_{mod}$=0.05 mV. The superconducting gap is 1.24 meV.
}
\end{figure}

\section{Summary}
We performed detailed STM/STS and QPI investigations of FeSe monolayer films grown on graphene/SiC(0001) substrate. The Fermi energy of FeSe monolayer is reduced to only a few milli-electron volts and can be tuned by graphene layers. Superconductor with ultra small Fermi pockets is a unique platform to study the exotic electron correlation effects\cite{PhysRevLett.97.107001}. The low Fermi energy implicates unconventional pairing mechanism. The retardation condition ($\omega_D\ll\epsilon_F$, where $\omega_D$ is the characteristic frequency of phonons) is crucial for the applicability of the conventional BCS theory. The violation of retardation condition in superconducting monolayer FeSe suggests pairing mechanism beyond BCS theory \cite{Lee16}. 
In general, the thin monolayer films of high temperature superconductors, both cuprates \cite{Jiang14,Zhang19} and iron-based superconductors \cite{Xue12}, have great potential to achieve deeper understanding of  high $T_{\rm{C}}$ superconductivity. 

Another attractive property of FeSe monolayer is that its exceedingly low Fermi energy is comparable to the superconducting gap.  The competition and cooperation of these energies may lead to new physics in the crossover regime. We propose FeSe monolayer as a distinctive system to study novel quantum states, such as BCS-BEC crossover. In the current experiments, the BCS-BEC crossover hasn't been realized yet. Further fine-tuning of the carrier density and correlation effect may lead to such many body states.
 
This work is supported by the National Natural Science Foundation of China (Grants No. 11934001 and No. 12074211) and the Ministry of Science and Technology of China (Grants No. 2016YFA0301002 and No. 2018YFA0305603).

\bibliography{PRB}
\bibliographystyle{apsrev4-2}

\end{document}